# Computing System Congestion Management Using Exponential Smoothing Forecasting

James F Brady
Capacity Planner for the State of Nevada
jfbrady@admin.nv.gov

*An overloaded computer must finish what it starts and not start what will fail or hang. A congestion management algorithm the author developed, and Siemens Corporation patented for telecom products, effectively manages traffic overload with its unique formulation of Exponential Smoothing forecasting. Siemens filed for exclusive rights to this technique in 2003 and obtained US patent US7301903B2 in 2007 with this author, an employee at the time of the filing, the sole inventor. A computer program, written in C language, which exercises the methodology is listed at the end of this document and available on GitHub.*

**1.0 Introduction**

There are countless examples of computing systems crashing or slowing to a virtual standstill due to an offered workload that exceeds available capacity. Application developers usually ignore the problem until disaster strikes and they are forced to address the issue. Strategies vary but are typically focused on one of three restrictive actions, capping queue length, throttling CPU usage, or limiting response time. When the chosen threshold is reached, action is taken to deny or delay new requests for service but permit transaction sequences underway to continue.

The approach presented here is based on limiting response time using measurements that are statistically smoothed with a uniquely formulated Exponential Smoothing forecasting model. Exponential Smoothing has not gained a great deal of acceptance by practitioners for this type of application because of some implementation issues thought to be insurmountable which are successfully addressed within this methodology.

Section 2 sets the stage with a web-based environment where overload control is implemented using exponentially smoothed response time measurements. Section 3 provides an introductory overview of the Exponential Smoothing forecasting model and compares its statistical properties to a more commonly used method. Issues associated with Exponential Smoothing as a prediction tool are discussed in Section 4 and adaptation of the complete overload control algorithm is described in Section 5. An overview of the computer program which exercises the algorithm is contained in Section 6, and some summary remarks are made in Section 7. The mathematics used throughout this discussion is algebra and the key reference, [BROW63], can be purchased online. As a convenience, links are setup in this document for internal as well as external references and section / figure locations. To return from internal links use the Alt + left arrow keys.

**2.0 Computing System Congestion Management Illustration**

Figure 1 is an example of a computing system congestion management setup that applies response time measurements. It shows a Web/App server and a Database server where round trip time is captured by the Web/App Server for a selected set of HTTP GET and POST events. These latency numbers are used for congestion control by statistically smoothing them to produce an "expected" response time which is compared to an overload threshold value. If the threshold is exceeded when the comparison is made, the next request for service is denied or delayed, but transaction sequences already in progress continue unabated to completion.

This is only one example of how smoothed response time can be used for congestion control. The implementation contained in US patent US7301903B2 is traffic flow control for a multimedia telecommunications system.

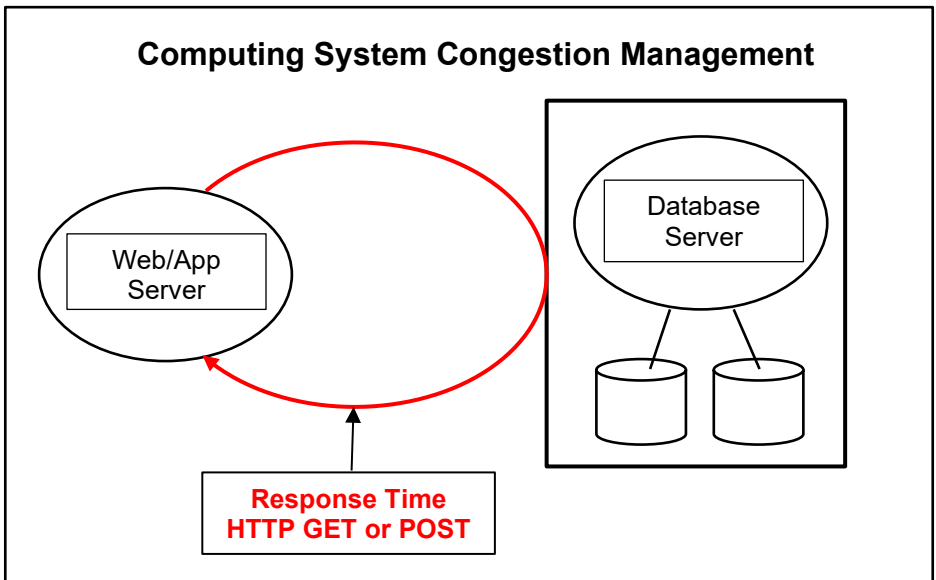

**Figure 1: Computing System Congestion Management Illustration**

The focus of this discussion is not a specific congestion management implementation, but the algorithm used to produce the smoothed response times needed for its operation. Exponential Smoothing time series forecasting is chosen as the basis of the smoothing algorithm because it requires a minimum amount of data to be retained and weighs recent data more heavily than older data. The next section contains an overview of Exponential Smoothing and compares its statistical properties to those of the more commonly used Moving Average.

### 3.0 Exponential Smoothing Vs Moving Average

Exponential Smoothing is a statistical forecasting technique that provides a time series forecast using the most recent data value and the previous forecast as input. Named for its data discounting properties, Exponential Smoothing lessens the weight given to a specific data point geometrically with the addition of subsequent observations to the time series. Although very small after many observations have occurred, the weight a data value contributes to the forecast is never zero. Eq. 1 is the most basic form of an exponentially smoothed forecast called Single Exponential Smoothing.

$$F_t = S_t(x) = \alpha x_t + (1 - \alpha)S_{t-1}(x) \qquad (1)$$

$Where$:

$F_t = forecast\ for\ point\ in\ time\ t$

$S_t(x) = first\ smoothed\ statistic\ at\ point\ in\ time\ t$

$\alpha = smoothing\ constant\ 0 < \alpha < 1$

$x_t = observation\ for\ point\ in\ time\ t$

This approach is contrasted with a Moving Average, Eq. 2, which produces its forecast by computing the arithmetic mean of the most recent N data values. This method weighs all N values in the calculation range equally and gives no weight to data outside that range.

$$M_t = \frac{x_t + x_{t-1} + \cdots + x_{t-N+1}}{N} \qquad (2)$$

$Where$:
$M_t = moving\ average\ forecast$

$N = number\ of\ moving\ average\ data\ points$

A Moving Average requires N data points to be stored for initiating and continuing the forecast while Exponential Smoothing needs only the most recent data point and the previous forecast. Figure 2 shows the difference in weight given to data for Exponential Smoothing Vs Moving Average. Each observation in the twenty point Moving Average gets an equal weight of $\frac{1}{N} = \frac{1}{20} = .05$, whereas, the most recent observation being Exponentially Smoothed gets a weight of $\alpha = .10$ and the oldest observation gets $\alpha(1 - \alpha)^{20-1} = .10(1 - .10)^{19} = .013509$.

$$\text{Exponential Smoothing } (S_t(x)) \text{ Vs Moving Average } (M_t)$$

$$\text{Weight Given } S_i(x) = \alpha, \alpha(1-\alpha), \cdots, \alpha(1-\alpha)^{i-1} \cdots$$

$$\text{Weight Given } M_i = \frac{1}{N}, \frac{1}{N}, \cdots, \frac{1}{N} \cdots$$

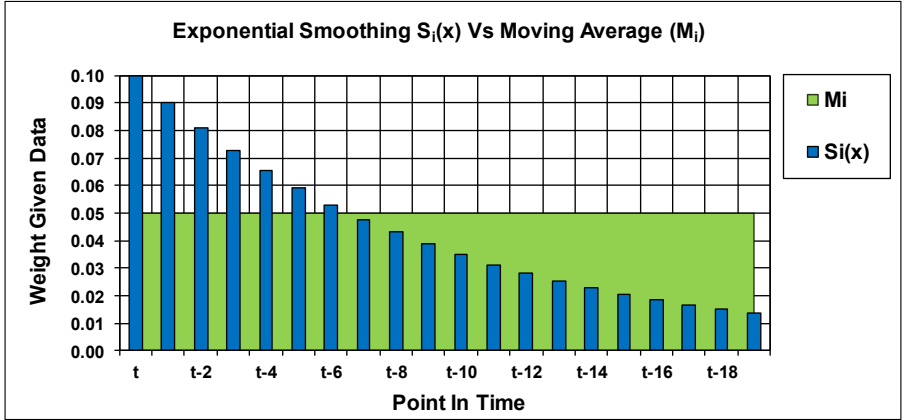

| | | α = | 0.10 |
|---|---|---|---|
| | | Weight Given | |
| i | Time | $S_i(x)$ | $M_i$ |
| 1 | t | 0.100000 | 0.050000 |
| 2 | t-1 | 0.090000 | 0.050000 |
| 3 | t-2 | 0.081000 | 0.050000 |
| 4 | t-3 | 0.072900 | 0.050000 |
| 5 | t-4 | 0.065610 | 0.050000 |
| 6 | t-5 | 0.059049 | 0.050000 |
| 7 | t-6 | 0.053144 | 0.050000 |
| 8 | t-7 | 0.047830 | 0.050000 |
| 9 | t-8 | 0.043047 | 0.050000 |
| 10 | t-9 | 0.038742 | 0.050000 |
| 11 | t-10 | 0.034868 | 0.050000 |
| 12 | t-11 | 0.031381 | 0.050000 |
| 13 | t-12 | 0.028243 | 0.050000 |
| 14 | t-13 | 0.025419 | 0.050000 |
| 15 | t-14 | 0.022877 | 0.050000 |
| 16 | t-15 | 0.020589 | 0.050000 |
| 17 | t-16 | 0.018530 | 0.050000 |
| 18 | t-17 | 0.016677 | 0.050000 |
| 19 | t-18 | 0.015009 | 0.050000 |
| 20 | t-19 | 0.013509 | 0.050000 |

**Figure 2: Exponential Smoothing Vs Moving Average Weight Given Time Series Samples**

For details concerning the mathematical formulation of both models and a discussion of their data discounting characteristics see [BROW63].

### 4.0 Exponential Smoothing Issues and Algorithm Summary

Many practitioners have found Exponential Smoothing to be an intuitively appealing approach to forecasting but were disappointed with the outcome. This author believes this dissatisfaction results from a less than adequate approach to two basic implementation issues.

1. Forecasting startup
2. Forecasting a ramp

The first issue results from the fact that significant weight is given to the initial forecast in Eq. 1, $S_0(x)$, for several subsequent forecasting periods. For example, the sum of the twenty weights in the Figure 2 **blue** column is .878423, leaving .121577 for the initial forecast, which is more than the .100000 given to the most recent data point. This problem is mitigated by restructuring Eq. 1 and applying that restructured formula at startup.

The second issue is important because Eq. 1 forecasts a ramp with a bias and response times ramp up rapidly as a system approaches a congested state.

### 4.1 Forecasting Startup

The significant weight given the initial forecast is demonstrated by expanding Eq. 1 over time as shown in Eq. 3 and detailed in Appendix A.

$$S_t(x) = \alpha x_t + \alpha(1-\alpha)x_{t-1} + \cdots + \alpha(1-\alpha)^{i-1}x_{t-i-1} + \cdots + (1-\alpha)^i S_0(x) \qquad (3)$$

The Eq. 3 expansion illustrates that after $i$ observations the weight given the initial estimate, $S_0(x)$, is $(1-\alpha)^i$, which implies that when $\alpha = .10$ and $i = 10$ the weight given $S_0(x) = .348678$. This tenth forecast weight is over three times the .100000 weight given the most recent observation, $x_t$, and even greater for forecasts $0 < i < 10$.

Figure 3 extends this example to twenty points in time where the weight given $S_0(x)$ is compared to the weight for the sum of the observations, $\sum S_i(x)$. The weight attributed to $\sum S_i(x)$ does not exceed that for $S_0(x)$ until the seventh observation, $i = 7$, where $\sum S_i(x) = .521703$ and $S_0(x) = .478297$.

$$\text{Weight Given } \sum S_i(x) = \alpha + \alpha(1-\alpha) + \cdots + \alpha(1-\alpha)^{i-1}$$

$$\text{Weight Given } S_0(x) = (1-\alpha), (1-\alpha)^2, \cdots, (1-\alpha)^i$$

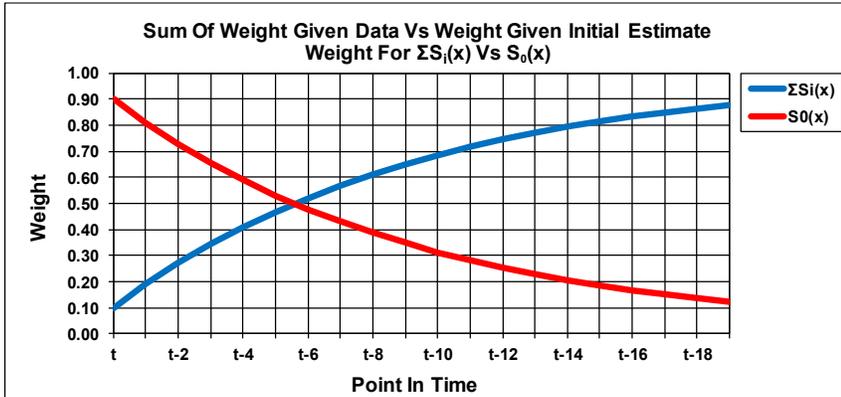

| α = | 0.10 | | |
|---|---|---|---|
| | | | Weight Given |
| i | Time | $S_i(x)$ | $\Sigma S_i(x)$ | $S_0(x)$ |
| 1 | t | 0.100000 | 0.100000 | 0.900000 |
| 2 | t-1 | 0.090000 | 0.190000 | 0.810000 |
| 3 | t-2 | 0.081000 | 0.271000 | 0.729000 |
| 4 | t-3 | 0.072900 | 0.343900 | 0.656100 |
| 5 | t-4 | 0.065610 | 0.409510 | 0.590490 |
| 6 | t-5 | 0.059049 | 0.468559 | 0.531441 |
| 7 | t-6 | 0.053144 | 0.521703 | 0.478297 |
| 8 | t-7 | 0.047830 | 0.569533 | 0.430467 |
| 9 | t-8 | 0.043047 | 0.612580 | 0.387420 |
| 10 | t-9 | 0.038742 | 0.651322 | 0.348678 |
| 11 | t-10 | 0.034868 | 0.686189 | 0.313811 |
| 12 | t-11 | 0.031381 | 0.717570 | 0.282430 |
| 13 | t-12 | 0.028243 | 0.745813 | 0.254187 |
| 14 | t-13 | 0.025419 | 0.771232 | 0.228768 |
| 15 | t-14 | 0.022877 | 0.794109 | 0.205891 |
| 16 | t-15 | 0.020589 | 0.814698 | 0.185302 |
| 17 | t-16 | 0.018530 | 0.833228 | 0.166772 |
| 18 | t-17 | 0.016677 | 0.849905 | 0.150095 |
| 19 | t-18 | 0.015009 | 0.864915 | 0.135085 |
| 20 | t-19 | 0.013509 | 0.878423 | 0.121577 |

**Figure 3: Sum of Weight Given Data Vs Weight Given Initial Estimate**

The solution to this problem is the forecasting startup formula, Eq. 4, with $n$ representing the observation sequence number. Eq. 4 is simply Eq. 1 with $\frac{1}{n} = \alpha$ when $n \leq \frac{1}{\alpha}$. The expansion of Eq. 4 below with $n = 1, 2, and\ 3$ shows that it is a recursive way to calculate the arithmetic mean for each of the first $n$ observations using only $n, x_t, and\ S_t(x)$, as input.

*Given*:

$n = observation\ sequence\ number$

*For*: $n \leq \frac{1}{\alpha}$

$$F_t = S_t(x) = \frac{1}{n}x_t + \left(1 - \frac{1}{n}\right)S_{t-1}(x) \qquad (4)$$

*When*:

$n = 1$:

$$S_1(x) = x_1$$

$n = 2$:

$$S_2(x) = \frac{1}{2}x_2 + \left(1 - \frac{1}{2}\right)x_1 = \frac{1}{2}x_2 + \frac{1}{2}x_1$$

$n = 3$:

$$S_3(x) = \frac{1}{3}x_3 + \left(1 - \frac{1}{3}\right)\frac{x_2 + x_1}{2} = \frac{1}{3}x_3 + \left(\frac{2}{3}\right)\frac{x_2 + x_1}{2} = \frac{1}{3}x_3 + \frac{1}{3}x_2 + \frac{1}{3}x_1$$

Eq. 4 eliminates the need to create an initial estimate for Eq. 1 because the initial forecast is the first observation, $x_1$. Figure 4 illustrates how Eq. 4 is implemented to reduce the impact of the initial estimate on subsequent forecasts. The **red** column, containing the $S_0(x)$ heading, reflects the traditional weight given the initial estimate based on $(1-\alpha)^i$, including the $10^{th} = .348678$. The **blue**/**green** column, with the $X_1$ heading, shows the first observation as the initial estimate, with its first three weights as enumerated above, 1, $\frac{1}{2}$, $\frac{1}{3}$. This Eq. 4 recursive mean calculation, **blue** weights, continues thru $n = \frac{1}{\alpha}$ or $X_{i=10} = \frac{1}{10}$. After that, Eq. 1 is applied, **green** weights, using this last Eq. 4 result as a starting point.

$$Weight\ Given\ Initial\ Estimate\ (S_0(x)\ Vs\ X_1)$$

$$Weight\ Given\ S_0(x) = (1-\alpha), (1-\alpha)^2, \cdots, (1-\alpha)^i$$

$$Weight\ Given\ X_1 = 1, \frac{1}{2}, \ldots, \frac{1}{i}, \ldots, \frac{1}{n}, \frac{(1-\alpha)}{n}, \ldots, \frac{(1-\alpha)^{i-n}}{n}$$

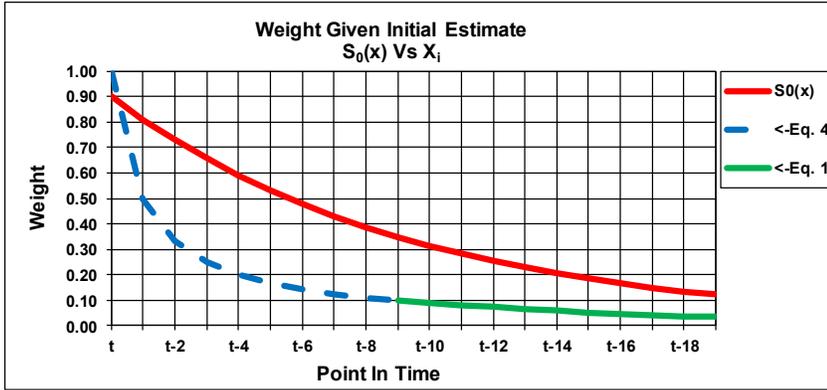

| | | α = | 0.10 | | |
|---|---|---|---|---|---|
| | | | Weight Given | | |
| i | Time | | $S_0(x)$ | $X_1$ | |
| 1 | t | | 0.900000 | 1.000000 | <-Eq. 4 |
| 2 | t-1 | | 0.810000 | 0.500000 | \| |
| 3 | t-2 | | 0.729000 | 0.333333 | \| |
| 4 | t-3 | | 0.656100 | 0.250000 | \| |
| 5 | t-4 | | 0.590490 | 0.200000 | \| |
| 6 | t-5 | | 0.531441 | 0.166667 | \| |
| 7 | t-6 | | 0.478297 | 0.142857 | \| |
| 8 | t-7 | | 0.430467 | 0.125000 | \| |
| 9 | t-8 | | 0.387420 | 0.111111 | \| |
| 10 | t-9 | | 0.348678 | 0.100000 | i = n |
| 11 | t-10 | | 0.313811 | 0.090000 | <-Eq. 1 |
| 12 | t-11 | | 0.282430 | 0.081000 | \| |
| 13 | t-12 | | 0.254187 | 0.072900 | \| |
| 14 | t-13 | | 0.228768 | 0.065610 | \| |
| 15 | t-14 | | 0.205891 | 0.059049 | \| |
| 16 | t-15 | | 0.185302 | 0.053144 | \| |
| 17 | t-16 | | 0.166772 | 0.047830 | \| |
| 18 | t-17 | | 0.150095 | 0.043047 | \| |
| 19 | t-18 | | 0.135085 | 0.038742 | \| |
| 20 | t-19 | | 0.121577 | 0.034868 | \| |

**Figure 4: Weight Given Initial Estimate ($S_0(x)\ Vs\ X_1$)**

Clearly, the **blue**/**green** line lessons the impact of the initial estimate on future forecasts by recursively averaging through the first $n$ data points. Note that the number of startup iterations Eq. 4 produces is inversely proportional to the value of the smoothing constant, a desirable property because initial estimates have greater impact when smoothing constants are small.

### 4.2 Forecasting a Ramp

Eq. 1 is a constant model, $F_t = a_t$, which forecasts step shifts in a random walk process without a bias. This response is shown in the Figure 5 top graph and corresponding **blue**/**red** "Step" data in its table. If, however, this same model is used to forecast a trend line or ramp, $F_t = a_t + b_t L$, it will track the ramp with a bias that eventually becomes $\frac{(1-\alpha)}{\alpha} b_t$. This ramp response is illustrated in the Figure 5 **blue**/**red** bottom graph and "Ramp" data. The bias is enumerated in the **light gray** column with the $\frac{(1-.2)}{.2} \times 10 = 40.0$ bias at the bottom of the table.

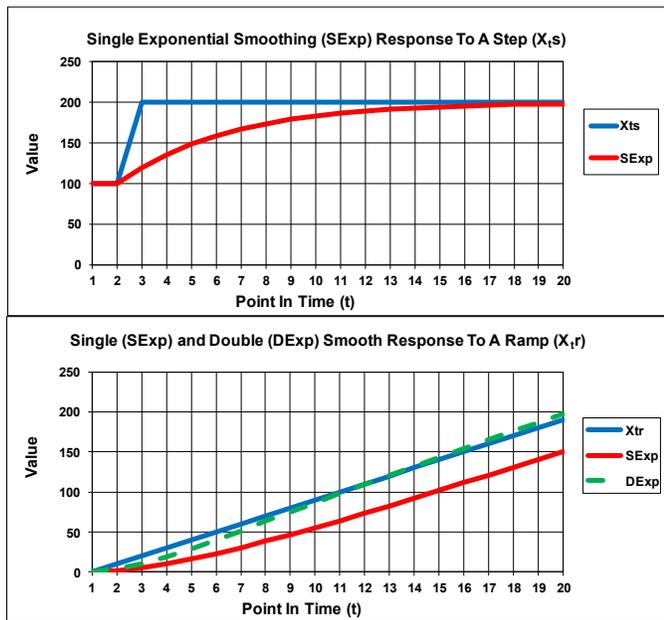

| α = | 0.20 | | | | | |
|---|---|---|---|---|---|---|
| | Step | | | Ramp | | |
| t | $X_t s$ | SExp | $X_t r$ | SExp | SExp Bias | DExp |
| 1 | 100 | 100.00 | 0 | 0.00 | 0.00 | 0.00 |
| 2 | 100 | 100.00 | 10 | 2.00 | 8.00 | 4.00 |
| 3 | 200 | 120.00 | 20 | 5.60 | 14.40 | 10.80 |
| 4 | 200 | 136.00 | 30 | 10.48 | 19.52 | 19.52 |
| 5 | 200 | 148.80 | 40 | 16.38 | 23.62 | 29.52 |
| 6 | 200 | 159.04 | 50 | 23.11 | 26.89 | 40.34 |
| 7 | 200 | 167.23 | 60 | 30.49 | 29.51 | 51.65 |
| 8 | 200 | 173.79 | 70 | 38.39 | 31.61 | 63.22 |
| 9 | 200 | 179.03 | 80 | 46.71 | 33.29 | 74.90 |
| 10 | 200 | 183.22 | 90 | 55.37 | 34.63 | 86.58 |
| 11 | 200 | 186.58 | 100 | 64.29 | 35.71 | 98.19 |
| 12 | 200 | 189.26 | 110 | 73.44 | 36.56 | 109.69 |
| 13 | 200 | 191.41 | 120 | 82.75 | 37.25 | 121.07 |
| 14 | 200 | 193.13 | 130 | 92.20 | 37.80 | 132.30 |
| 15 | 200 | 194.50 | 140 | 101.76 | 38.24 | 143.40 |
| 16 | 200 | 195.60 | 150 | 111.41 | 38.59 | 154.37 |
| 17 | 200 | 196.48 | 160 | 121.13 | 38.87 | 165.21 |
| 18 | 200 | 197.19 | 170 | 130.90 | 39.10 | 175.95 |
| 19 | 200 | 197.75 | 180 | 140.72 | 39.28 | 186.58 |
| 20 | 200 | 198.20 | 190 | 150.58 | 39.42 | 197.12 |
| | | | | Bias = | 40.00 | |

**Figure 5: Exponential Smoothing Response to Step and Ramp**

The solution is to apply Double Exponential Smoothing as shown in Eq. 5 with $x_t$ in Eq. 1 replaced by $S_t(x)$.

$$S_t^{[2]}(x) = \alpha S_t(x) + (1 - \alpha) S_{t-1}^{[2]}(x) \tag{5}$$

$Where$:

$S_t^{[2]}(x) = second\ smoothed\ statistic\ at\ point\ in\ time\ t$

The Double Exponential Smoothing forecast, Figure 5 **green** dotted line and table column, is produced with Eq. 6 by applying a linear combination of Eq. 1 and Eq. 5 as shown in Eq. 8 and Eq. 9.

$$F_t = a_t + b_t\, L \tag{6}$$

$Where$:
$$L = 1 \tag{7}$$

$$a_t = 2S_t(x) - S_t^{[2]}(x) \tag{8}$$

$$b_t = \frac{\alpha}{(1-\alpha)} \left[ S_t(x) - S_t^{[2]}(x) \right] \tag{9}$$

A derivation of these expressions is contained in [BROW63]. Higher order terms, such as a quadratic component, can be added but they tend to increase volatility in the forecast.

### 4.3 Algorithm Summary

Eq. 4 addresses issue 1 by eliminating the dominant weight given to the initial forecast and assigning equal weight to the first $n$ observations incrementally. Eq. 6 through Eq. 9 resolve issue 2 by implementing Double Exponential Smoothing on an ongoing basis, thus tracking a ramp without a bias.

Eq. 4 is used at startup to produce the response time forecast when $n \leq \left\lfloor \frac{1}{\alpha} \right\rfloor$ and the Double Exponential Smoothing equations are applied when generating ongoing forecasts. Eq. 10 is implemented at Double Exponential Smoothing startup, $n = \left\lfloor \frac{1}{\alpha} \right\rfloor$, making $S_t(x)$ the forecast.

$$S_{t_{n=\lfloor \frac{1}{\alpha} \rfloor}}^{[2]}(x) = S_{t_{n=\lfloor \frac{1}{\alpha} \rfloor}}(x) \tag{10}$$

The only data items stored ongoing in this Double Exponential Smoothing setup are $n, S_t(x), and\ S_t^{[2]}(x)$.

## 5.0 Algorithm Implementation

There are two implementation issues to consider when adapting the formulas just discussed to a real time forecasting environment like the one shown in Figure 1. First, there is a potential requirement to create an integer math formulation of the expressions and second, a mechanism needs to be in place to reset the algorithm based on the time that has elapsed since the last update.

An integer math formulation is likely necessary if the algorithm is implemented in the operating system as part of a device driver. Linux, as an example, does not reliably support floating point math in the kernel. The second issue focuses on resetting the forecasting algorithm if it has been a long time since an observation has taken place.

### 5.1 Integer Math Formulation

The integer math conversion is produced by substituting $\frac{1}{n}$ for $\alpha$ where $0 < \alpha \leq .5$ into the equations previously introduced and rearranging. The results of these manipulations are shown below with Eq. 12 as an integer replacement for Eq. 4 and Eq. 13 thru Eq. 19 integer replacements for Eq. 1 and Eq. 5 thru Eq. 10, ongoing.

The integer formulation halves the range of $\alpha$ because an $\alpha > .5$ makes $n_\alpha = 1$ and $b_t$ in Eq. 16 undefined. Eq. 12 and Eq. 16 through Eq. 18 perform multiplication, addition, and subtraction before division based on standard C language operator precedence. The $\lfloor integer\ floor \rfloor$ bars are used sparingly below to avoid unnecessary clutter. Appendix B provides a detailed mapping of these integer equations to their non-integer counterparts.

*For*: $0 < \alpha \leq .5$

*Let*:
$$n_\alpha = \left\lfloor \frac{1}{\alpha} \right\rfloor \quad (11)$$

*Startup*: $n \leq n_\alpha$
$$F_t = S_t(x) = \frac{x_t + (n-1)S_{t-1}(x)}{n} \quad (12)$$

*Ongoing*:
$$F_t = a_t + b_t L \quad (13)$$

*Where*:
$$L = 1 \quad (14)$$

$$a_t = 2S_t(x) - S_t^{[2]}(x) \quad (15)$$

$$b_t = \frac{S_t(x) - S_t^{[2]}(x)}{(n_\alpha - 1)} \quad (16)$$

$$S_t(x) = \frac{x_t + (n_\alpha - 1)S_{t-1}(x)}{n_\alpha} \quad (17)$$

$$S_t^{[2]} = \frac{S_t(x) + (n_\alpha - 1)S_{t-1}^{[2]}(x)}{n_\alpha} \quad (18)$$

*Initial Condition*:
$$S_{t_{n=n_\alpha}}^{[2]}(x) = S_{t_{n=n_\alpha}}(x) \quad (19)$$

Although this integer formulation halves the range of the smoothing constant, there are few forecasting situations where over 50% of the forecasting weight should be given to the current observation. One way to minimize the truncation error incurred using integer data types is to choose a smoothing constant with an integer inverse, e.g., $n_{\alpha=.1} = \left\lfloor \frac{1}{.1} \right\rfloor = 10$ and $n_{\alpha=.2} = \left\lfloor \frac{1}{.2} \right\rfloor = 5$. The default setting in the Appendix C computer program is $n_\alpha = 10$.

These equations are used to produce the results contained in Figure 6 for $n_\alpha = 10$ using the input data and C language computer program in Appendix C.

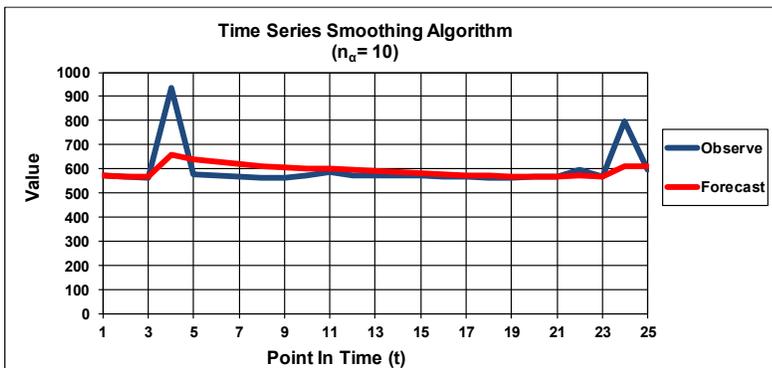

*Time Series Smoothing Algorithm Integer Formulation*

$$n_\alpha = \left\lfloor \frac{1}{.1} \right\rfloor = 10$$

| Time Series Smoothing Algorithm | | | | | | | |
|---|---|---|---|---|---|---|---|
| $n_\alpha =$ | 10 | | | | | | |
| t | Observe | Forecast | n | $S_t(x)$ | $S_t^{[2]}(x)$ | $a_t$ | $b_t$ |
| 1 | 571 | 571 | 1 | 571 | 571 | 571 | 0 |
| 2 | 565 | 568 | 2 | 568 | 568 | 568 | 0 |
| 3 | 564 | 566 | 3 | 566 | 566 | 566 | 0 |
| 4 | 936 | 658 | 4 | 658 | 658 | 658 | 0 |
| 5 | 576 | 641 | 5 | 641 | 641 | 641 | 0 |
| 6 | 574 | 629 | 6 | 629 | 629 | 629 | 0 |
| 7 | 569 | 620 | 7 | 620 | 620 | 620 | 0 |
| 8 | 563 | 612 | 8 | 612 | 612 | 612 | 0 |
| 9 | 562 | 606 | 9 | 606 | 606 | 606 | 0 |
| 10 | 570 | 602 | 10 | 602 | 602 | 602 | 0 |
| 11 | 585 | 599 | 10 | 600 | 601 | 599 | 0 |
| 12 | 573 | 594 | 10 | 597 | 600 | 594 | 0 |
| 13 | 570 | 589 | 10 | 594 | 599 | 589 | 0 |
| 14 | 574 | 586 | 10 | 592 | 598 | 586 | 0 |
| 15 | 570 | 581 | 10 | 589 | 597 | 581 | 0 |
| 16 | 567 | 576 | 10 | 586 | 595 | 577 | -1 |
| 17 | 567 | 574 | 10 | 584 | 593 | 575 | -1 |
| 18 | 563 | 570 | 10 | 581 | 591 | 571 | -1 |
| 19 | 562 | 568 | 10 | 579 | 589 | 569 | -1 |
| 20 | 569 | 568 | 10 | 578 | 587 | 569 | -1 |
| 21 | 569 | 567 | 10 | 577 | 586 | 568 | -1 |
| 22 | 595 | 571 | 10 | 578 | 585 | 571 | 0 |
| 23 | 566 | 568 | 10 | 576 | 584 | 568 | 0 |
| 24 | 796 | 612 | 10 | 598 | 585 | 611 | 1 |
| 25 | 594 | 609 | 10 | 597 | 586 | 608 | 1 |

**Figure 6: Time Series Smoothing Algorithm Integer Formulation**

This is the same input data included in US patent US7301903B2. The first 10 **red** forecast values are the arithmetic mean, Eq. 12, but after that, $t > 10$, the Double Exponential Smoothing portion of the algorithm, Eq. 13, takes control of the forecast. The integer setup is preferred by this author because it is fast, not resource intensive, and can be implemented at the operating system kernel level.

## 5.2 Forecast Reset

Because response times are event based, their sampling rate increases with higher traffic volume, allowing for quick reaction to an overload condition. Contrast this type of metric with an interval-based measurement like CPU busy which requires multiple fixed time interval samples to be taken that are a minimum of one second in duration. CPU busy also suffers from the fact that it is a resource busy indicator, not a direct measure of system congestion level.

Response times are a system congestion barometer only if the time between samples is small because, if it has been a long time since the last observation, e.g., five seconds, the smoothed response time is stale and the statistically smoothed forecast, $F_t$, needs to be reset. The most glaring example of this situation occurs in the laboratory when overload traffic is being offered to a system and is suddenly stopped and restarted later below the overload level. The congestion management controls that are set will remain in effect for a few event sequences until the forecasting algorithm reacts to the smaller response times being produced by the reduced traffic rate.

The solution to this problem is to reset the forecasting algorithm based on the time since the last update. This is accomplished by storing the wall clock time of the last update and if it has been at least, the example, five seconds since this last measurement, the current observation number is set to one, $n = 1$, from its ongoing value, $n_\alpha$. This procedure resets the forecasting mechanism to the startup state, Eq. 12, making the current forecast equal to the value of the current observation, $x_t$. Since it has been a long time since the last event has occurred, this event's response time will be small or, if it is not, a system failure of some kind has likely taken place. The Figure 7 ramp model illustrates this mechanism when $n_\alpha = 5$ and the reset time delay begins at $t = 11$.

$$Time\ Series\ Smoothing\ Algorithm\ With\ Reset$$
$$Integer\ Formulation$$
$$(began\ reset\ delay\ at\ count = 11)$$

$$n_\alpha = \left\lceil \frac{1}{.2} \right\rceil = 5$$

| Time Series Smoothing Algorithm With Reset | | | | | | | |
|---|---|---|---|---|---|---|---|
| $n_\alpha =$ | 5 | reset_c = | 11 | | | | |
| t | Observe | Forecast | n | $S_t(x)$ | $S_t^{[2]}(x)$ | $a_t$ | $b_t$ |
| 1 | 0 | 0 | 1 | 0 | 0 | 0 | 0 |
| 2 | 10 | 5 | 2 | 5 | 5 | 5 | 0 |
| 3 | 20 | 10 | 3 | 10 | 10 | 10 | 0 |
| 4 | 30 | 15 | 4 | 15 | 15 | 15 | 0 |
| 5 | 40 | 20 | 5 | 20 | 20 | 20 | 0 |
| 6 | 50 | 32 | 5 | 26 | 21 | 31 | 1 |
| 7 | 60 | 43 | 5 | 32 | 23 | 41 | 2 |
| 8 | 70 | 55 | 5 | 39 | 26 | 52 | 3 |
| 9 | 80 | 68 | 5 | 47 | 30 | 64 | 4 |
| 10 | 90 | 80 | 5 | 55 | 35 | 75 | 5 |
| 11 | 100 | 94 | 5 | 64 | 40 | 88 | 6 |
| 12 | 110 | 110 | 1 | 110 | 110 | 110 | 0 |
| 13 | 120 | 115 | 2 | 115 | 115 | 115 | 0 |
| 14 | 130 | 120 | 3 | 120 | 120 | 120 | 0 |
| 15 | 140 | 125 | 4 | 125 | 125 | 125 | 0 |
| 16 | 150 | 130 | 5 | 130 | 130 | 130 | 0 |
| 17 | 160 | 142 | 5 | 136 | 131 | 141 | 1 |
| 18 | 170 | 153 | 5 | 142 | 133 | 151 | 2 |
| 19 | 180 | 165 | 5 | 149 | 136 | 162 | 3 |
| 20 | 190 | 178 | 5 | 157 | 140 | 174 | 4 |
| 21 | 200 | 190 | 5 | 165 | 145 | 185 | 5 |
| 22 | 210 | 204 | 5 | 174 | 150 | 198 | 6 |
| 23 | 220 | 216 | 5 | 183 | 156 | 210 | 6 |
| 24 | 230 | 228 | 5 | 192 | 163 | 221 | 7 |
| 25 | 240 | 239 | 5 | 201 | 170 | 232 | 7 |

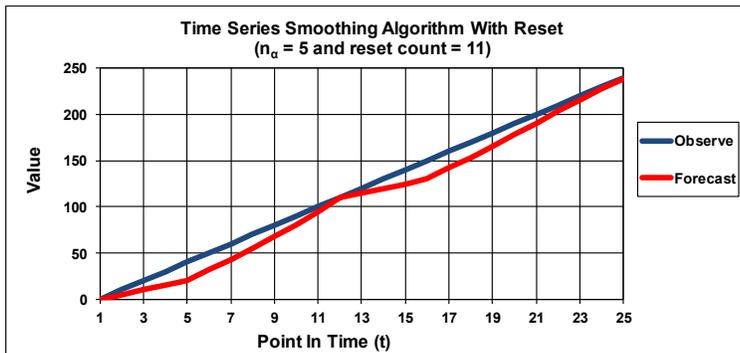

**Figure 7: Time Series Smoothing Algorithm Integer Formulation with Reset**

This reset example illustrates that during the averaging periods, $t = 1 - 5\ and\ t = 12 - 16$, the **forecasts** are increasing at a much lower rate than the **observed** values but when Double Exponential Smoothing kicks in, $t = 6 - 10\ and\ t = 17 - 25$, the forecast asymptotically approaches the observations. Of course, the forecast exactly matches the observation at $t = 12$ since it is the first event after a 6 second pause occurred with a 5 second reset interval.

## 6.0 Computer Program

The equations and diagrams in the previous three sections explain the algorithm but may not be as helpful to the practitioner as the computer program contained in Appendix C. This C language program, which implements the equations in Section 5.1 and the forecast reset features of Section 5.2, can be copied, compiled, and run. It can also be downloaded from GitHub.

Appendix C has four parts:
1. Program Input
2. Program Output
3. Program Compile
4. Program Source Listing

To create and run the program:
1. Highlight and paste the program input data into a file named – time_series_smooth_input.txt
2. Highlight and past the program source listing into a file named – time_series_smooth.c
3. Compile the source program:
    a. gcc time_series_smooth.c -Wall -o time_series_smooth.exe
4. Execute the compiled program
    a. time_series_smooth.exe time_series_smooth_input.txt

Options:
1. -h = help
2. -n = n_alpha - integer value of [1/alpha] default is 10
3. -r = reset smoother at count value plus one
4. -t = reset smoother time interval default is 5 seconds
5. -w = write verbose output to comma delimited file

The function in the program that contains the full algorithm is "time_series_smooth()". This function's source code, call in the main program, prototype, and data structure (EXP_SMOOTH_DATA) are shown in **red**. The tab settings in the source listing are lost when pasting into a file but the program will compile and execute.

## 7.0 Summary

Managing congestion in a computing system is a difficult challenge that is usually ignored until an overload crashes the system or leaves it in an indeterminant state. This paper describes an approach to the problem that was applied in a telecommunication setting and worked so well the corporation who developed the technique filed for and obtained U.S. patent rights. Section 2 illustrates how this methodology can be adapted to a computing system environment using response time measurements between a Web/App server and Database server. This computing system example provides implementation context for the primary focus of the discussion, the forecasting algorithm.

The algorithm applies Exponential Smoothing forecasting to an event driven time series of response time measurements, statistically smoothing them, and comparing the result to an overload threshold. If the threshold is exceeded, new requests for service, e.g., login, are denied or delayed, but transaction sequences already started progress to completion. The forecasting startup approach in Section 4.1 as well as the forecast reset procedure in Section 5.2 solve some of problems users have encountered with Exponential Smoothing in the past. The integer formulation, Section 5.1, permits the algorithm's implementation within an operating system level device driver.

The mathematical formulations and associated illustrations contained in Section 3 thru Section 5 define the algorithm as well as show how and why it works. A thorough understanding of these equations is unnecessary because a computer program that exercises them is provided in Appendix C. This C language program, which implements all aspects of the methodology, can be highlighted, pasted into a file, compiled, and run or simply downloaded from GitHub.

The primary focus of US7301903B2 is telecommunications products, not computing systems, but if there are patent infringement concerns, the initial filing took place on November 4, 2003 and the grant normally ends 20 years after that filing. There is an extension of 848 day for this patent which makes March 1, 2026 the expiration date.

Although the focus of this paper is congestion management, the algorithm is a generic time series forecaster adaptable to a broad range of prediction situations. Implementation within a computing system monitoring or planning tool to statistically smooth resource consumption or response time service level measurements is one possibility.

## References

[BRAD07] J. F. Brady, Method and Apparatus for Achieving an Optimal Response Time in A Telecommunications System, United State Patent and Trademark Office, Patten No. US7301903B2, 2007.

[BROW63] R. G. Brown, Smoothing, Forecasting and Prediction of Discrete Time Series, Prentice-Hall, Inc., Englewood Cliffs, N.J., 1963.

[GIFF71] W. C. Giffin, Introduction to Operations Engineering, Irwin, Inc, Homewood Illinois, 1971.

## Copyrights and Trademarks

# Appendix A
# Single Exponential Smoothing Over Time

This appendix shows how Eq. 3 in Section 4.1 is derived by iterating the Eq. 1 Single Exponential Smoothing formula through multiple points in time, $t$. **Red** is used below to highlight the $t-1$ and $t-2$ expressions.

$Eq. 1$:
$$F_t = S_t(x) = \alpha x_t + (1-\alpha)S_{t-1}(x) \tag{1A}$$

$Iterating\ thru\ t-2$:
$$S_t(x) = \alpha x_t + (1-\alpha)S_{t-1}(x)$$
$$S_{t-1}(x) = \color{red}{\alpha x_{t-1} + (1-\alpha)S_{t-2}(x)}$$
$$S_{t-2}(x) = \color{red}{\alpha x_{t-2} + (1-\alpha)S_{t-3}(x)}$$
$$\Downarrow$$

$Substituting\ and\ expanding$:
$$S_t(x) = \alpha x_t + (1-\alpha)[\color{red}{\alpha x_{t-1} + (1-\alpha)S_{t-2}(x)}]$$
$$S_t(x) = \alpha x_t + \alpha(1-\alpha)x_{t-1} + (1-\alpha)^2 S_{t-2}(x)$$
$$S_t(x) = \alpha x_t + \alpha(1-\alpha)x_{t-1} + (1-\alpha)^2[\color{red}{\alpha x_{t-2} + (1-\alpha)S_{t-3}(x)}]$$
$$S_t(x) = \alpha x_t + \alpha(1-\alpha)x_{t-1} + \alpha(1-\alpha)^2 x_{t-2} + (1-\alpha)^3 S_{t-3}(x)$$
$$\vdots$$

$Eq. 3$:
$$S_t(x) = \alpha x_t + \alpha(1-\alpha)x_{t-1} + \cdots + \alpha(1-\alpha)^{i-1}x_{t-i-1} + \cdots + (1-\alpha)^i S_0(x) \tag{3A}$$

# Appendix B
# Derivation of the Integer Formulation of the Smoothing Algorithm from the Non-Integer Expressions

This appendix maps the non-integer Exponential Smoothing equations, Eq. 1 and Eq. 4 thru Eq. 10, to their integer counterparts, Eq. 12 thru Eq. 19. Expressions that are the same in both cases are excluded. The $\lfloor integer\ floor \rfloor$ bars are used sparingly in the integer expressions below but each component in those expressions is assumed to be implemented as an integer in a C language program using that language's standard operator precedence where multiplication, addition, and subtraction are performed before division.

*For*: $0 < \alpha \le .5$

*Let*:
$$n_\alpha = \left\lfloor \frac{1}{\alpha} \right\rfloor \quad (11B)$$

*Then*:
$$\frac{\alpha}{(1-\alpha)} \le \frac{\frac{1}{n_\alpha}}{(1-\frac{1}{n_\alpha})} = \frac{\frac{1}{n_\alpha}}{(\frac{n_\alpha-1}{n_\alpha})} = \frac{1}{(n_\alpha-1)} \quad (11iB)$$

$F_t$: *Eq*. 4 *to Eq*. 12
$$F_t = S_t(x) = \frac{1}{n}x_t + \left(1-\frac{1}{n}\right)S_{t-1}(x) \quad (4B)$$
$$F_t = S_t(x) = \frac{1}{n}x_t + \frac{(n-1)}{n}S_{t-1}(x) = \frac{x_t + (n-1)S_{t-1}(x)}{n} \quad (12B)$$

$b_t$: *Eq*. 9 *to Eq*. 16
$$b_t = \frac{\alpha}{(1-\alpha)}\left[S_t(x) - S_t^{[2]}(x)\right] \quad (9B)$$
$$b_t = \frac{1}{(n_\alpha-1)}\left[S_t(x) - S_t^{[2]}(x)\right] = \frac{\left[S_t(x) - S_t^{[2]}(x)\right]}{(n_\alpha-1)} \quad (16B)$$

$S_t(x)$: *Eq*. 1 *to Eq*. 17
$$F_t = S_t(x) = \alpha x_t + (1-\alpha)S_{t-1}(x) \quad (1B)$$
$$S_t(x) = \frac{1}{n_\alpha}x_t + \left(1-\frac{1}{n_\alpha}\right)S_{t-1}(x) = \frac{1}{n_\alpha}x_t + \left(\frac{n_\alpha-1}{n_\alpha}\right)S_{t-1}(x) = \frac{x_t+(n_\alpha-1)S_{t-1}(x)}{n_\alpha} \quad (17B)$$

$S_t^{[2]}(x)$: *Eq*. 5 *to Eq*. 18
$$S_t^{[2]}(x) = \alpha S_t(x) + (1-\alpha)S_{t-1}S_t^{[2]}(x) \quad (5B)$$
$$S_t^{[2]}(x) = \frac{1}{n_\alpha}S_t(x) + \left(1-\frac{1}{n_\alpha}\right)S_{t-1}^{[2]}(x) = \frac{1}{n_\alpha}S_t(x) + \left(\frac{n_\alpha-1}{n_\alpha}\right)S_{t-1}^{[2]}(x) = \frac{S_t(x)+(n_\alpha-1)S_{t-1}^{[2]}(x)}{n_\alpha} \quad (18B)$$

# Appendix C
# Computer Program in C Language

This appendix contains the C language computer program outlined in Section 6. The "Program Input" and "Program Source" Listing" can each be highlighted and pasted into a file. When the program is compiled and run the results in "Program Output" are produced. The smoothing algorithm is the "time_series_smooth()" function with its source statements, call in the main program, prototype, and data structure (EXP_SMOOTH_DATA) highlighted in **red**.

## Program Input

```
time_series_smooth_input.txt
1 571
2 565
3 564
4 936
5 576
6 574
7 569
8 563
9 562
10 570
11 585
12 573
13 570
14 574
15 570
16 567
17 567
18 563
19 562
20 569
21 569
22 595
23 566
24 796
25 594
```

## Program Output

```
> time_series_smooth.exe time_series_smooth_input.txt

---------Time Series Smoothing Algorithm----------
n_alpha = 10   reset_time = 5
     count___observe__forecast______diff___diffsum
         1        571       571          0         0
         2        565       568         -3        -3
         3        564       566         -2        -5
         4        936       658        278       273
         5        576       641        -65       208
         6        574       629        -55       153
         7        569       620        -51       102
         8        563       612        -49        53
         9        562       606        -44         9
        10        570       602        -32       -23
        11        585       599        -14       -37
        12        573       594        -21       -58
        13        570       589        -19       -77
        14        574       586        -12       -89
        15        570       581        -11      -100
        16        567       576         -9      -109
        17        567       574         -7      -116
        18        563       570         -7      -123
```

```
        19       562       568        -6      -129
        20       569       568         1      -128
        21       569       567         2      -126
        22       595       571        24      -102
        23       566       568        -2      -104
        24       796       612       184        80
        25       594       609       -15        65
```

## Program Compile

```
gcc time_series_smooth.c -Wall -o time_series_smooth.exe
```

## Program Source Listing

C language program, "time_series_smooth.c", which exercises the Section 5.1 algorithm.

```c
/***************************************************************************
 * This is a C language source listing of a computer program which exercises
 * a time series smoothing algorithm based on Exponential Smoothing.
 *
 * The time_series_smooth function contains the smoothing algorithm.
 *
 * Author: James F. Brady 2019
 ***************************************************************************
 * Definitions:
 * alpha = smoothing constant 0 < alpha < 1
 * n_alpha = set integer value of [1/alpha]
 * N_ALPHA = default integer value of [1/alpha]
 * reset_time = set sample number reset time
 * RESET_TIME = default sample number reset time
 * reset_count = reset smoother at count value plus one
 * xt = current sample (observation)
 * stx1 = first smoothed statistic
 * stx2 = second smoothed statistic
 * n = sample number (may be reset)
 * ft = forecast
 * last_update_time = timestamp of last update
 * count = running sample count for output
 * diff = xt-ft for output
 * diffsum = sum of xt-ft for output
 ***************************************************************************
 * Args:
 *    input file name
 * Options:
 *    -h = help
 *    -n = n_alpha - integer value of [1/alpha] default is 10
 *    -r = reset smoother at count value plus one
 *    -t = reset smoother time interval default is 5 seconds
 *    -w = write verbose output to comma delimited file
 ***************************************************************************/
/**********************************
 * Includes
 **********************************/
#include <sys/time.h>
#include <stdio.h>
#include <time.h>
#include <stdlib.h>
#include <stdbool.h>
#include <string.h>
#include <unistd.h>
#include <stdint.h>
```

```c
/*********************************
 * Defines
 *********************************/
#define N_ALPHA 10
#define RESET_TIME 5

/*********************************
 * Data Structures
 *********************************/
typedef struct
{
    int     n_alpha;
    int     stx1;
    int     stx2;
    int     n;
    int     ft;
    int     reset_time;
    int     last_update_time;
}EXP_SMOOTH_DATA;

/*********************************
 * Prototypes
 *********************************/
int time_series_smooth(EXP_SMOOTH_DATA *, int);
int sleep(int);

/*********************************
 * Main
 *********************************/
int main(int argc, char **argv)
{
  FILE            *in_file;
  FILE            *out_file;
  int             opt;
  int             count;
  int             xt;
  int             diff;
  bool            errorFlag = false;
  int             diffsum = 0;
  int             reset_count = 0;
  char            out_file_name[100];
  int             out_file_set = 0;
  EXP_SMOOTH_DATA data;

  /********************
   * Initialize data
   ********************/
  memset((char *)&data, 0, sizeof(data));
  /**************************
   * Initialize variables
   **************************/
  data.n_alpha = N_ALPHA;
  data.reset_time = RESET_TIME;
  /*********************************
   * Get options
   *********************************/
  while ((opt = getopt(argc, argv, "hn:r:t:w:")) != EOF)
  {
    switch (opt)
```

```c
{
  /***********************************
   * help
   ***********************************/
  case 'h':
    printf("\n**************************************************************\n");
    printf("* Args:\n");
    printf("* input file name\n");
    printf("*\n");
    printf("* Options:\n");
    printf("* -h = help\n");
    printf("* -n = n_alpha - integer value of [1/alpha] default is 10\n");
    printf("* -r = reset smoother at count value plus one\n");
    printf("* -t = reset smoother time interval default is 5 seconds\n");
    printf("* -w = write verbose output to comma delimited file\n");
    printf("**************************************************************\n");
    errorFlag = true;
    break;
  /*****************************************************
   * Option -n  integer value of [1 / alpha]
   *****************************************************/
  case 'n':
    if ((data.n_alpha = strtol(optarg, 0, 0)) == 0)
    {
      fprintf(stderr,"Invalid n_alpha = %s\n",optarg);
      errorFlag = true;
    }
    if (data.n_alpha < 0)
    {
      fprintf(stderr,"Invalid n_alpha = %d\n",data.n_alpha);
      errorFlag = true;
    }
    break;
  /*****************************************************
   * Option -r reset smoother at count value plus one
   *****************************************************/
  case 'r':
    if ((reset_count = strtol(optarg, 0, 0)) == 0)
    {
      fprintf(stderr,"Invalid reset_count = %s\n",optarg);
      errorFlag = true;
    }
    if (reset_count < 0)
    {
      fprintf(stderr,"Invalid reset_count = %d\n",reset_count);
      errorFlag = true;
    }
    break;
  /*****************************************************
   * Option -t reset smoother time interval
   *****************************************************/
  case 't':
    if ((data.reset_time = strtol(optarg, 0, 0)) == 0)
    {
      fprintf(stderr,"Invalid reset_time = %s\n",optarg);
      errorFlag = true;
    }
    if (data.reset_time < 0)
    {
      fprintf(stderr,"Invalid reset_time = %d\n",data.reset_time);
```

```c
          errorFlag = true;
        }
        break;
      /*************************************************
       * Option -w write output to comma delimited file
       *************************************************/
      case 'w':
        if (strcpy(out_file_name,optarg) != 0)
        {
          out_file_set = 1;
        }
        else
          {
            fprintf(stderr,"Invalid file name = %s\n",optarg);
            errorFlag = true;
          }
        break;
      /**********************************
       * default
       **********************************/
      default:
        break;
    }
  }
  /*******************************************
   * The errorFlag is set
   *******************************************/
  if (errorFlag)
  {
    exit(1);
  }
  /***********************************
   * Check args passed
   ***********************************/
  if (argc - optind < 1)
  {
    fprintf(stderr,"usage: %s [opt-hn:r:t:w:] file name\n",argv[0]);
    exit(1);
  }
  /*********************
   * Open input file
   *********************/
  if ((in_file = fopen(argv[argc-1],"r")) == NULL)
  {
    fprintf(stderr,"Error opening input file = %s\n", argv[argc-1]);
    exit(1);
  }
  /*********************
   * Print heading
   *********************/
  fprintf(stdout,"\n");
  fprintf(stdout,"---------Time Series Smoothing Algorithm---------\n");
  fprintf(stdout,"n_alpha = %d",data.n_alpha);
  fprintf(stdout,"   reset_time = %d",data.reset_time);
  if (reset_count)
  {
    fprintf(stdout,"   reset_count = %d",reset_count);
  }
  fprintf(stdout,"\n      count    observe   forecast       diff    diffsum\n");
  /*********************************************
```

```c
 * Option -w open output file and write heading
 **************************************************/
if (out_file_set)
{
  if ((out_file = fopen(out_file_name,"w")) == NULL)
  {
    if (out_file == NULL)
    {
      fprintf(stderr,"Error opening output file = %s\n",out_file_name);
      exit(1);
    }
  }
  fprintf(out_file,"%s\n","Time Series Smoothing Algorithm");
  fprintf(out_file,"%s%d","n_alpha = ,",data.n_alpha);
  fprintf(out_file,"%s%d",",,reset_t = ,",data.reset_time);
  if (reset_count)
  {
    fprintf(out_file,"%s%d",",,reset_c = ,",reset_count);
  }
  fprintf(out_file,"\n%s\n","count,observe,forecast,diff,diffsum,n,stx1,stx2");
}

/***************************************************************
 * Read input file and write xt and ft to stdout
 ***************************************************************/
for (;;)
{
  /*****************************average
   * Read input file
   *****************************/
  if (fscanf(in_file, "%d%d", &count, &xt) == EOF)
  {
    break;
  }
  /********************************************************
   * Call time series smoothing function
   ********************************************************/
  time_series_smooth(&data, xt);

  /********************************************************
   * Write xt and ft and diffs to stdout
   ********************************************************/
  diff = xt - data.ft;
  diffsum += diff;
  fprintf(stdout,"%10d%10d%10d%10d%10d\n",
                  count,
                  xt,
                  data.ft,
                  diff,
                  diffsum);
  /********************************************************
   * Option -w write xt and ft and diffs to output file
   ********************************************************/
  if (out_file_set)
  {
    fprintf(out_file,"%d%s%d%s%d%s%d%s%d%s%d%s%d%s%d\n",
            count,",",xt,",",data.ft,",",diff,",",diffsum,
                  ",",data.n,",",data.stx1,",",data.stx2);
  }
  /***********************************************
```

```c
       * Option -r reset the smoother at count value
       **********************************************/
      if (reset_count && reset_count == count)
      {
        sleep(data.reset_time + 1);
      }
  }
  /********************
   * Close input File
   *******************/
  fclose(in_file);
  /*****************************
   * Option -w close output file
   *****************************/
  if (out_file_set)
  {
    fclose(out_file);
  }
  return(0);
}

/******************************************************
 * Time Series Smoothing Algorithm
 ******************************************************/
int time_series_smooth(EXP_SMOOTH_DATA *data, int xt)
{
  struct timeval curr_time;

  /*********************************
   * Check xt range
   *********************************/
  if (xt > INT32_MAX / data->n_alpha)
  {
    xt = INT32_MAX / data->n_alpha;
  }
  if (xt < INT32_MIN / data->n_alpha)
  {
    xt = INT32_MIN / data->n_alpha;
  }

  /*************************************
   * Get the current time
   *************************************/
  gettimeofday(&curr_time, NULL);

  /***********************************************************
   * Reset data->n if last update > RESET_TIME_INTERVAL
   ***********************************************************/
  if (((curr_time.tv_sec - data->last_update_time)) >
        data->reset_time)
  {
    data->n = 0;
  }

  /**************************************
   * Update data->last_update_time
   **************************************/
  data->last_update_time = curr_time.tv_sec;

  /***************************************************************
```

```c
 * Double Exp Smooth if data->n >= data->n_alpha
 ****************************************************************/
if (data->n >= data->n_alpha)
{
  data->stx1 = (xt + (data->n_alpha-1) * data->stx1)
               / data->n_alpha;
  data->stx2 = (data->stx1 + (data->n_alpha-1) * data->stx2)
               / data->n_alpha;
  /*************************************************************
   * If data->n_alpha > 1
   **************************************************************/
  if (data->n_alpha > 1)
  {
    data->ft = 2 * data->stx1 - data->stx2 + (data->stx1 - data->stx2)
               / (data->n_alpha-1);
  }
  /*************************************************************
   * If data->n_alpha = 1
   **************************************************************/
  else
  {
    data->ft = data->stx1;
  }
}
/*********************************************************************
 * Average if data->n < data->n_alpha
 **********************************************************************/
else
{
  data->n++;
  data->stx1 = (xt + (data->n-1) * data->stx1)
               / data->n;
  data->stx2 = data->stx1;
  data->ft = data->stx1;
}
return(0);
}
```